\documentstyle[11pt]{article}
\newcommand{\zero}{\setcounter{equation}{0}}

\begin{document}

\thispagestyle{empty}

\begin{center}

{\Large \bf Gravitational field around a time-like current-carrying screwed cosmic string in
scalar-tensor theories}

\vspace{0.3 true cm}

V. B. Bezerra$^*$\footnote{valdir@fisica.ufpb.br},
C.N. Ferreira$^{\dagger \ddagger}$\footnote{crisnfer@cbpf.br},
J. B. Fonseca-Neto$^*$ and A.A. R. Sobreira$^*$

\vspace{.5 true cm}

$^*$Departamento de F\'{\i}sica, Universidade Federal da Para\'{\i}ba,

Caixa Postal 5008, 58059-970, Jo\~ao Pessoa, PB, Brazil

$^\dagger $Instituto de
F\'{\i}sica, Universidade Federal do Rio de Janeiro

Caixa Postal 68528, 21945-910, Rio de Janeiro, RJ, Brazil

$^\ddagger $Grupo de F\'{\i}sica
Te\'orica Jos\'e Leite Lopes, Petr\'opolis, RJ, Brazil

\end{center}


\begin{abstract}

In this paper we obtain the space-time generated by a time-like current-carrying
superconducting screwed cosmic string(TCSCS). This gravitational field is obtained
in a modified scalar-tensor theory in the sense that torsion is taken into account. 
We show that this solution is comptible with a torsion field generated by the scalar 
field $\phi $. The analysis of gravitational effects of a TCSCS shows up that the 
torsion effects that appear in the physical frame of Jordan-Fierz can be described 
in a geometric form given  by contorsion term plus a symmetric part which contains 
the scalar gradient. As an important application of this solution, we consider the 
linear perturbation method developed by Zel'dovich, investigate the accretion of cold 
dark matter due to the formation of wakes when a TCSCS moves with speed $v$ and discuss 
the role played by torsion. Our results are compared with those obtained for cosmic strings 
in the framework of scalar-tensor theories without taking torsion into account.

\end{abstract}

\newpage

\section{Introduction }

Scalar tensor-theories of gravity represent the simplest and natural generalization
of Einstein's theory of general relativity. The earliest scalar-tensor theories considered
a massless scalar field with constant coupling to matter\cite{Jordan}. Later, scalar-tensor theories
were generalized by having a scalar field self-interaction and dynamical coupling
to matter\cite{Bergmann}. More recently, these theories have been generalized further
to the case of multiple scalar fields\cite{Esposito}. In these theories\cite{Shapiro1,Gaspperini}, 
the gravitational interaction is mediated by one or several long range fields in addition to the usual
tensor field of Einstein's theory of general relativity. The principal consequence of these theories 
is the fact that at sufficiently high energy scales\cite{Hehl}-\cite{Kim} they can be
relevant. If gravity is essentially a scalar-tensor theory, there will be direct implications for 
cosmology and experimental tests of the gravitational interaction\cite{will94,will93,dam}.

Thus, it seems natural to investigate a general theory of gravity which involves scalar-tensor fields, 
especially those aspects connected with the gravitational fields generated by topological defects
such as cosmic strings\cite{Vilenkin} and study their gravitational and
cosmological consequences. The gravitational field of a cosmic string, in the context of
the theory of general relativity is quite remarkable: a particle placed at rest around
a straight, infinite, static cosmic string will not be attracted to it: there is no local
gravity. Its space-time is locally flat but not globally. Thus, the external gravitational
field due to a cosmic string may be approximatelly  described by what is commonly called conical
geometry. The non-trivial topology of this space-time leads to a number of interesting effects,
such as, for example, gravitational lensing\cite{Gott}, emission of radiation by a freely
moving particle\cite{Aliev}, existence of an electrostatic self-force\cite{Blinet} on an electric
charged particle at rest, and the so-called gravitational Aharonov-Bohm effect\cite{Ford}, among
others. Therefore, the richness of effects produced by this defect as well as the new ideas this 
object brought to general relativity seems to justify the interest in the study of this structure, 
and specifically the possible role played by it in the framework of cosmology due to the fact that 
it carries a large energy density, especially in the context of scalar-tensor theories of gravity 
due to their relevance at high energies in which scale topological defects could be formed.
In this context, some authors have studied solutions for cosmic strings in Brans-Dicke\cite{rom},
in dilaton theory \cite{greg} and in situations with more general scalar-tensor couplings\cite{mexg,Cris}. 

Torsion fields play an important role in the geometry of a cosmic string whose presence could have been
influenced the formation and evolution of  structures in the Universe \cite{William}. As an example of 
effects produced by torsion we can mention the one which corresponds to the contribution to neutrino 
oscillations \cite{Adak}. The fact that torsion could influence some physical phenomena led several authors to argue
that torsion may have been an important element in the early universe, when quantum effects of
gravity were drastically important\cite{Sabbata1,Yishi} and for this reason have to be taken into account.
In the context of cosmology torsion is important and produces modifications of kinematic quantities, like shear,
vorticity, acceleration and expansion\cite{Palle}, \cite{Trautman}-\cite{Ellis}.

Taking into account the arguments concerning the importance of torsion and that scalar-tensor theories of
gravity could be important at least at sufficiently high energy in which scenario topological defects, like
a cosmic string could be formed, we investigated a modified scalar-tensor theory of gravity in the sense
that a torsion field is present\cite{VBCF,VBHMCF}. Based on this modified scalar-tensor theory we determined
the space-time generated by a screwed superconducting bosonic cosmic string and studied some of its features.

Following this approach, in this paper we will determine the space-time generated by a time-like
current-carrying screwed cosmic string(TCSCS) and investigate some of its physical consequences.
In this background, an interesting aspect to consider are the anisotropies of the Cosmic Microwave Background
Radiation (CMBR). These anisotropies were analysed using numerical simulations of high-resolution which demonstrated
the existence of scale densities and revealed the presence of significant small-scale structure
on strings or ''wiggliness''\cite{Bennett88}-\cite{Albrecht89}. This substructure has
important dynamical consequences and also causes loop formation on scales much smaller than
the horizon. These results suggest that long strings are more important than loops
in seeding density perturbations \cite{Silk84,Stebbins89}.

The assumption that a string network\cite{Pogosian1,Pogosian2} presents wiggles, leads to the
conclusion that wiggles cosmic strings accompanying strings
with low velocities, present a significant peak in the CMBR and an enhancement
in the matter power spectrum. The scale of the wiggles is much smaller than the characteristic length
of the string. In fact it was argued that a distant observer will not be able to resolve the details
of the wiggly structure, and therefore, in this case it is not possible to distinguish the wiggle cosmic
string of those generated in the presence of torsion. For this reason it is important to investigate with
details this framework with torsion and its possible effects.

The interesting effect which can be studied is the accretion of the cold dark matter by wakes
when the cosmic string is moving in a space-time with torsion in the context of a scalar-tensor
theory. This effect was studied taking into account the string power spectrum  obtained in simulations,
projecting this forward to the present day using linear theory of transfer functions for
both cold and hot dark matter. This was an invaluable first step, but further developments are necessary
because strings create nonlinear objects at early times and power spectrum provides an
incomplete description of non-Gaussian perturbations. Our work assumes the same idea, but using torsion as
an essential element. In the study of wiggles, the approach considered frequently is
the one proposed by Zel'dovich\cite{Zel}. Here, we will assume that this is the correct mechanism to study
wiggles also when torsion is present. This assumption is justified by the fact that this mechanism
considers nonlinear effects. We will determine the time-dependent metric in linearized gravity for
arbitrary evolving string configurations. As we will see the use of the Zel'dovich approximation in this
scenario induces perturbations grows in a cold dark matter universe.
In doing so, we postulate that the small-scale structures existing in wiggles strings can be approximately scaled
by the geometrical deformation that torsion produces \cite{VBCF}.

In this work we study the implications on these phenomena when we have a TCSCS.
Our purposes are to obtain the gravitational field surrounding a TCSCS and study some of its
consequences, in particular, how the cosmological effects of long strings are affected by torsion
and scalar fields as compared with the corresponding results in general relativity. One important
effect which will be studied is the accretion of the cold dark matter by wakes when
a TCSCS is moving. Thus, in order to study this phenomenon we will analyse the formation and evolution
of wakes in this space-time with special emphasis to the role played by torsion in the process of
formation of wakes. Also we will consider a possible explanation to the anisotropies of the CMBR,
which will be done by the analysis of perturbations in this background. This problem
is most conveniently studied using the linear perturbation methods developed by
Zel'dovich \cite{Zel}. In this approach we consider the universe in the matter-dominated
era $t> t_{eq}$ with scale factor $a(t) \sim t^{2/3}$, and average density given by
$\rho_{av} = 1/6\pi G t^2 $.

This paper is organized as follows: In Section II we present a short description of scalar-tensor
theories with torsion fields with a discussion of geodesics based on energy conservation and in terms of
contorsion. In Section III we obtain the solution that corresponds
to a screwed cosmic string carrying a time-like current(TCSCS) by applying a method used by
Linet\cite{Linet} to solve the linearized Einstein's equations. In Section IV, we study the scattering
by a TCSCS, and in Section V, the Zel'dovich approximation in a theory with torsion is
introduced. In Section VI, we study the accretion of cold dark matter by wakes. Finally, in Section VII
we provide some closing remarks.

\section{Setting up scalar-tensor theories of gravity with torsion \zero}

\noindent

In this Section, we consider some basic features of scalar-tensor theories of gravity with the
inclusion of torsion. An important aspect to point out in these theories is the information concerning
the presence of torsion which comes out in the geodesic equations written in the Jordan-Fierz frame.
It is important to call attention to the fact that usually test particles are affected by torsion if
they are fermions. In our model, indeed, bosonic particles can also be affected by the torsion background.
This effect appear in the geodesic equations which present contributions arising from torsion. If we only
consider metric aspects, that is, if the metricity condition  $\nabla_{\mu}g_{\alpha \beta}=0$
is assumed, we find that the connection of a Riemann-Cartan manifold $U_4$, is given by

 \begin{equation}
\Gamma_{\lambda \nu}^{\hspace{.3 true cm} \alpha} =
{\{^{\alpha}_{\lambda \nu}\}}_{JF} +
 K_{\lambda \nu}^{\hspace{.3 true cm}\alpha} \; , \label{kont1}
\end{equation}

\noindent
where ${\{^{\alpha}_{\lambda \nu}\}_{JF}} $ is the usual Christoffel symbol evaluated in the
Jordan-Fierz frame with metric, $\tilde g_{\mu \nu}$. The contortion tensor,
$K_{\lambda \nu}^{\hspace{.3 true cm}\alpha}$, reads

\begin{equation}
K_{\lambda \nu}^{\hspace{.3 true cm}\alpha} =
-\frac{1}{2}(S_{\lambda \hspace{.2 true cm} \nu}^{\hspace{.1 true
cm} \alpha} + S_{\nu \hspace{.2 true cm} \lambda}^{\hspace{.1 true
cm} \alpha} - S_{\lambda \nu}^{\hspace{.3 true cm} \alpha}) \; .\label{kont2}
\end{equation}

\noindent
where $S_{\lambda \nu}^{\hspace{.3 true cm} \alpha}$ is the torsion.
As torsion can also exist in absence of fermionic particles\cite{Kim}, let us assume that
the dilaton can generate torsion which can be written, in terms of this field, as

\begin{equation}
S_{\mu\nu}^{\hspace{.3 true cm} \lambda} =
( \delta^{\lambda}_{\mu} \partial_{\nu} \tilde \phi -
\delta^{\lambda}_{\nu} \partial_{\mu}\tilde \phi)/
2\tilde \phi. \label{torsion}
\end{equation}

In this case the curvature tensor is defined as in a Riemannian space, using however
the connections defined on $U_4$, instead of the Christoffel symbols. The action is the same of
the Refs. \cite{Kim,Gaspperini,VBCF} and takes the form

\begin{equation}
I=\frac 1{16\pi }\int d^4x\sqrt{{-\tilde g }}\left[ {\tilde \phi}{ \tilde R(\{\}) }-
\frac{\kappa (\tilde \phi)}{\tilde \phi}\partial _\mu \tilde  \phi
\partial^\mu \tilde \phi \right] +  I_m ( \tilde g_{\mu \nu}, \Psi_m )  \label{acao1}\; ,
\end{equation}

\noindent
where $\kappa (\tilde \phi) = \omega(\tilde \phi) - \epsilon $, with $\omega(\tilde \phi) $ being a general
function of the scalar field and $\epsilon $ is the torsion coupling constant\cite{Gaspperini}.
The scalar curvature $\tilde R(\{\})$ is evaluated in the Jordan-Fierz frame and the action coresponding
to the matter fields is represented by $I_m ( \tilde g_{\mu \nu}, \Psi_m ) $.

In the physical frame of Jordan-Fierz, the equations for the metric, $\tilde g_{\mu \nu}$, are

\begin{equation}
\tilde G_{\mu\nu} =  \frac{\kappa (\tilde \phi) }{\tilde \phi^2}
\left(\partial_{\mu}\tilde \phi\partial_{\nu} \tilde \phi -
\frac{1}{2} \tilde g_{\mu\nu} \tilde
g^{\alpha\beta}\partial_{\alpha}\tilde \phi \partial_{\beta}
\tilde \phi \right)+ 8\pi G \tilde T_{\mu\nu}\label{eins0},
\end{equation}

If we consider that the action of matter does not have fermionic
fields, we find from Bianchi identities that

\begin{equation}
\partial_{\nu}\left(\sqrt{-\tilde g} \tilde T^{\mu \nu}\right) + \sqrt{-\tilde g}{
\{^{\mu}_{\nu \alpha} \}_{JF}} \tilde T^{\mu \nu} =0.\label{geod1}
\end{equation}

This result permits us to write the geodesic equations in terms of the connection only and therefore,
the contortion term does not appear. Then, in this case, the geodesic equations are

\begin{equation}
\frac{d^2x^{\mu }}{d\tau^2} + \{^{\mu}_{\alpha  \beta } \}_{JF}  \frac{dx^{\alpha}}{d\tau}
\frac{dx^{\beta}}{d\tau}  =0, \label{geode1}
\end{equation}

This is the correct result corresponding to geodesic equations in the cases where spin is not present.
It is worth calling attention to the fact that this result does not means that torsion in our work does
not contribute to geodesics. In fact, all contributions arising from torsion, in the case where spin is not
present, are coded in the metric written in Jordan-Fierz frame and taking into account the contortion.
The torsion contribution to geodesic equations becomes evident if we write these equations as a function of the
dilaton field. The action proposed in Eq.(\ref{acao1}), in Jordan-Fierz frame can be transformed
to another frame called Einstein frame, using a field dependent conformal transformation \cite{Niew} given by

\begin{equation}
\tilde{g}_{\mu\nu} = \Lambda^2(\phi)g_{\mu\nu},
\end{equation}
and defining the quantities $\Lambda^2(\phi)$ and $\alpha^2(\phi)$ as

\begin{equation}
\begin{array}{ll}
G\Lambda^2(\phi) = \tilde{\phi}^{-1}\\
\alpha^2(\phi) \equiv \left( \frac{\partial \ln \Lambda(\phi)}{\partial
\phi} \right)^2 = [2\omega(\tilde{\phi}) + 3]^{-1},
\end{array}
\label{conform}
\end{equation}

Therefore, we can calculate all quantities in the Einstein frame and reexpress all them in the physical 
frame of Jordan-Fierz just by using the above conformal transformation.

Now, let us analyse Eq.(\ref{geode1}) with the torsion written explicitly. To do this, it would be
better to write Eq.(\ref{geode1}) as a function of the contorsion. Thus, let us consider the Christoffel
symbols in the Jordan-Fierz frame as a sum of the Christoffel symbols in the Einstein frame
$\{^{\mu}_{\alpha  \beta } \} $, the contortion given by (\ref{kont2}) and the dilaton $\phi $, as

\begin{equation}
\{^{\mu}_{\alpha  \beta } \}_{JF}  = \{^{\mu}_{\alpha  \beta } \} + K_{(\alpha \beta)}^{\hspace{ .4 true cm} \mu }
+ \frac{\alpha(\phi)}{2} \left( \delta^{\mu}_{\alpha }\partial_{\beta} \phi +
\delta^{\mu}_{\beta}\partial_{\alpha} \phi \right) \label{Chisjf},
\end{equation}
where the contorsion, $K_{(\alpha \beta)}^{\hspace{ .4 true cm} \mu}$, can be written, explicitly, as

\begin{equation}
K_{(\alpha \beta)}^{\hspace{ .4 true cm} \mu} = \frac{\alpha(\phi)}{2} \left( \delta^{\mu}_{\alpha }\partial_{\beta} 
\phi + \delta^{\mu}_{\beta}\partial_{\alpha} \phi - 2 g_{\alpha \beta} g^{\mu \nu} \partial_{\nu} \phi
\right),\label{contorsion1}
\end{equation}

\noindent
and the dilaton $\phi$ is the solution of the equation of motion

\begin{equation}
\Box _g\phi =-4\pi G\alpha (\phi )T  \label{eq1},
\end{equation}

\noindent
where
\begin{equation}
\Box_g \phi = \frac{1}{\sqrt{-g}}\partial_{\mu}
\left[\sqrt{-g}\kappa(\phi) \partial^{\mu} \phi\right],
\end{equation}
is the d'Alembertian in this background and $\kappa( \phi) $ is defined as

\begin{equation}
\kappa(\phi) = 1- 2 \epsilon \alpha^2(\phi) \label{kappa},
\end{equation}
with $\epsilon$ being the torsion coupling constant.

The important point to call attention here is the fact that the symmetric part of the contorsion
appears in the Christoffel symbols in the Jordan-Fierz frame, which is the frame where the physical
quantities are mensurables. Notice that the torsion effects are taken into account in (\ref{Chisjf}) and this is a
peculiarity of the Jordan-Friez frame. As we will see in Section IV, the expression for the Newtonian force on a
test particle put this point into evidence.

\section{ Time-like current-carrying screwed cosmic string in scalar-tensor theories}

In this Section we will consider the appropriate action for matter fields, $I_m$, which can be
used to obtain the solution of a time-like current-carrying screwed cosmic string in scalar-tensor theories.
The model which we will consider here has been already discussed recently\cite{VBCF} in the magnetic 
case(space-like current). In order to recall some features of this model we will include a brief discussion 
in this Section, with the appropriate changes to be considered in the electric case.

Using the transformation given by (\ref{conform}), we can express action (\ref{acao1}) in the Einstein
frame in the following form

\begin{equation}
\begin{array}{lll}
I &=& \frac{1}{16\pi G}\int d^4x \sqrt{-g} \left[ R(\{\}) -
2\kappa (\phi) g^{\mu\nu}\partial_{\mu}\phi\partial_{\nu}\phi
\right]\\
&&\\
&+&\int d^4x\sqrt{-g}\left[ -\frac{1}{2}\Lambda^2\left(D_{\mu}\Phi
(D^{\mu}\Phi)^* - \frac{1}{2}D_{\mu} \Sigma (D^{\mu}
\Sigma )^*\right)\right] \\
&&\\
& -&\left. \frac{1}{16\pi}\left(F_{\mu \nu}F^{\mu \nu} + H_{\mu
\nu}H^{\mu \nu}\right) - \Lambda^2 V(|\Phi |, |\Sigma |) \right],
\end{array}
,\label{acao3}
\end{equation}

\noindent
where $D_{\mu} \Sigma =(\partial_{\mu} + ieA_{\mu}) \Sigma$ and $D_{\mu} \Phi=(\partial_{\mu} + iqB_{\mu}) \Phi$ are
the covariant derivatives, with $A_{\mu}$ and $B_{\mu}$ being the gauge fields and $\Phi$ and $\Sigma$ the scalar
fields. The field strengths are defined as usual as $F_{\mu \nu}=\partial_{\mu} A_{\nu} -
\partial_{\nu} A_{\mu}$ and $H_{\mu \nu}=\partial_{\mu} B_{\nu} - \partial_{\nu} B_{\mu}$,
Note that $\kappa (\phi)$, already defined, contains informations coming from the scalar-tensor
term, $\alpha^2(\phi)$, and from the torsion through the coupling constant, $\epsilon$.  In this work
we will consider high orders in $\kappa(\phi)$, because we are interested in the torsion contribution
at high energy scales, when possibly cosmic strings were formed.

The vortex configuration associated with the fields $(\Phi, B_{\mu})$ is given by

\begin{equation} \begin{array}{ll} \Phi = \varphi(r )e^{i\theta},\\
B_{\mu} = \frac{1}{q}[P(r) - 1]\delta^{\theta}_{\mu},
\end{array}\label{vortex1} \end{equation}

\noindent
with $(t,r,\theta,z)$ being the usual cylindrical coordinates with $r \geq 0$ and $0 \leq \theta < 2 \pi$.
The fields $\varphi(r) $ and $P(r)$ obey the same boundary conditions as the ordinary cosmic
strings\cite{Nielsen}, namely $\varphi(r) = \eta $ and $P(r) =0$ outside the string  and $\varphi(r) =0$ and $P(r)=1$
in the core. The electromagnetic properties are represented by the fields $(\Sigma , A_{\mu})$
with the configurations

\begin{equation}
\Sigma = \sigma(r)e^{i\zeta(z,t)}\label{vortex2},
\end{equation}

\begin{equation}
A_{\mu} = \frac{1}{e}[A_t(r) - \frac{\partial \zeta(z,t)}{\partial
t}]\delta_{\mu}^{t},
\label{config3}
\end{equation}

In the string core, the $\Sigma $-field acquires an expectation value and is responsible for the
time-like current carried by the gauge field $A_{\mu}$ which does not vanishes outside the string.
The potential $V(\varphi, \sigma)$ triggering the spontaneous symmetry breaking can be written in the most general
case as

\begin{equation} V(\varphi, \sigma) = \frac{\lambda_{\varphi}}{4} (
\varphi ^2 - \eta^2)^2 + f_{\varphi \sigma}\varphi ^2\sigma ^2 +
\frac{\lambda_{\sigma}}{4}\sigma ^4 -
\frac{m_{\sigma}^2}{2}\sigma^2 ,
\end{equation}

\vspace{.5 true cm}

\noindent
where $\lambda_{\varphi}$, $\lambda_{\sigma}$, $f_{\varphi \sigma}$ and
$m_{\sigma}$ are coupling constants. Considering the analogy with the ordinary
cosmic string case, this potential possesses all the ingredients necessary
to drive the formation of a screwed cosmic string with a time-like current.

Now, let us consider a cosmic string in a cylindrical coordinate system, in which situation we can write the
metric for the electric case in the Einstein frame, as\cite{PP, MacC}

\begin{equation}
ds^2 = e^{2(\gamma - \psi)}(dr^2 + dz^2 ) + \beta^2
e^{-2\psi}d\theta^2 - e^{2\psi}dt^2 \label{metric1},
\end{equation}

\noindent
where $\gamma, \psi$ and  $\beta$ depend only on $r$.

A straightforward calculation shows that the Einstein equations appropriately modified to taken into account
contributions coming from the scalar-tensor features and torsion, can be written, in the Einstein frame as

\begin{eqnarray}
R_{\mu \nu }&=&2\kappa( \phi) \partial _\mu \phi \partial
_\nu \phi +8\pi G(T_{\mu \nu} - \frac{1}{2}g_{\mu
\nu}T),\label{escalar1}\\
G_{\mu\nu} & = & 2\kappa( \phi) \partial_{\mu}\phi\partial_{\nu}\phi -
\kappa( \phi) g_{\mu\nu}g^{\alpha\beta}\partial_{\alpha}\phi\partial_{\beta}
\phi + 8\pi G T_{\mu\nu}\label{eins}.
\end{eqnarray}

Thus, the Einstein equations in the space-time given by Eq.(\ref{metric1}) reads

$$
\beta'' = 8 \pi G \beta ( T^t_t + T^r_r) e^{2(\gamma -
\psi)}
$$

\begin{equation}
(\beta \gamma')' = 8 \pi G \beta ( T^r_r +
T^{\theta}_{\theta}) e^{2(\gamma - \psi)},\label{eqq1}
\end{equation}

$$
(\beta \psi')'= 4 \pi G \beta ( T^t_t +
T^r_r + T^{\theta}_{\theta} -T^z_z)e^{2(\gamma -
\psi)}.
$$

The equation describing the scalar field $\phi $, in this background, is given by

\begin{equation}
(\beta
\kappa(\phi) \phi')'= 4\pi G \beta T \alpha(\phi) e^{2(\gamma -
\psi)}.\label{efi1}
\end{equation}

In order to solve Eqs.(\ref{eqq1}) and (\ref{efi1}), let us write explicitly the components of the
energy-momentum tensor in this case, which are given as
\vspace{.5 true cm}

\begin{eqnarray}
T^{t}_{t} & = &  - \frac{1}{2}\Lambda^{2}(\phi) \{
e^{2( \psi - \gamma)}(\phi'^{2} + \sigma'^{2}) + \frac{e^{2 \psi}}
{\beta^2}\phi^2P^2 + e^{-2 \psi}\sigma^2A_t^2  \nonumber \\
& & + \Lambda^{-2}(\phi) e^{-2\gamma}(\frac{A_t'^2}{4\pi e^2}) +
\Lambda^{-2}(\phi)\frac{e^{2(2 \psi - \gamma)}}{\beta^2}(\frac{P'^2}
{4\pi q^2}) + 2\Lambda^2(\phi)V(\phi,\sigma) \},  \\
&&\nonumber\\
&& \nonumber \\
T^{z}_{z} & = & - \frac{1}{2}\Lambda^2(\phi) \{ e^{2( \psi - \gamma)}
(\phi'^2 + \sigma^2) + \frac{e^2 \psi}{\beta^2}\phi^2P^2 - e^{-2 \psi}\sigma^2
A_t^2 \nonumber \\
& & - \Lambda^{-2}(\phi)e^{-2\gamma}(\frac{A_t'^2}{4\pi e^2}) + \Lambda^{-2}(\phi)
\frac{e^{2(2 \psi -\gamma)}}{\beta^2}(\frac{P'^2}{4 \pi q^2}) +  2\Lambda^2(\phi)
V(\phi,\sigma)  \},\\
&&\nonumber \\
&&\nonumber \\
T^{r}_{r} & = &  \frac{1}{2}\Lambda^2(\phi) \{ e^{2(\psi - \gamma)}
(\phi'^2 + \sigma'^2) - \frac{e^{2\psi}}{\beta^2} \phi^2P^2 +
e^{-2 \psi}\sigma^2A_t^2 \nonumber \\
& & - \Lambda^{-2}(\phi)e^{-2\gamma}(\frac{A_t'^2}{4\pi e^2}) + \Lambda^{-2}(\phi)
\frac{e^{2(2 \psi - \gamma)}}{\beta^2}(\frac{P'^2}{4\pi q^2}) - 2\Lambda^2(\phi)
V(\phi,\sigma) \}, \\
&&\nonumber \\
&&\nonumber \\
T^{\theta}_{\theta} & =  & - \frac{1}{2}\Lambda^2(\phi) \{ e^{2(\psi -
\gamma)}(\phi'^2 + \sigma'^2) - \frac{e^{2 \psi}}{\beta^2} \phi^2P^2 +
e^{-2 \psi}\sigma^2A_t^2 \nonumber \\
& & - \Lambda^{-2}(\phi)e^{-2\gamma}(\frac{A_t'^2}{4\pi e^2}) - \Lambda^{-2}(\phi)
\frac{e^{2(2 \psi - \gamma)}}{\beta^2}(\frac{P'^2}{4\pi q^2}) + 2\Lambda^2(\phi)
V(\varphi,\sigma) \}.
\end{eqnarray}

Note that in the present case of a time-like current, the $T^{r}_{r}$ and
$T^{\theta}_{\theta}$ components have different sign in the electromagnetic part as
compared to the space-like case\cite{VBCF}. In this situation, because only the temporal
component of the electromagnetic field is different from zero, the electric charge in
the core of the string does not vanish. In next sections we will investigate the
consequences of this fact, in the framework of the weak field approximation.

Now, we solve the previous set of equations given by (\ref{eqq1}), in the region outside
the TCSCS, that is, for $ r_0 \leq r \leq \infty $, where $r_0$ is the radius of the string.
In this region, the contribution to the energy-momentum tensor of the string reads

\begin{equation}
\begin{array}{llll}
T^{t}_{ t} =  - \frac{1}{2}e^{-2\gamma}(\frac{A_t'^2}{4\pi e^2}) &
T^{z}_{ z} =   \frac{1}{2}e^{-2\gamma}(\frac{A_t'^2}{4\pi e^2}) &
T^{r}_{ r} =  - \frac{1}{2}e^{-2\gamma}(\frac{A_t'^2}{4\pi e^2}) &
T^{t}_{ t} =   \frac{1}{2}e^{-2\gamma}(\frac{A_t'^2}{4\pi e^2})
\end{array}
\end{equation}

If we consider the asymptotic conditions, we can conclude that only the field $A_\mu$ does not vanish 
outside the string. In this case, the external solutions of Eq.(\ref{eqq1}) are formally the same
of the scalar-tensor theory \cite{Cris}, but the $\phi$-solution is different and comes from
the equation

\begin{equation}
\phi'= \kappa^{-1}(\phi) \frac{\lambda }{r}.
\end{equation}

The solutions for $\beta (r) $ and $\gamma(r) $ are the same obtained recently\cite{Cris} and are
written  as

\begin{equation}
\beta =Br, \hspace{1 true cm}
\gamma = m^2 \ln{r/r_0}.\label{eq2}
\end{equation}
where $B$ and $m$ are integration constants. If we use Brans-Dicke theory to estimate the order
of magnitude of the correction induced by $ \kappa^{-1}(\phi)\lambda $, considering
particular values of the parameter $\omega$ consistent with solar system experiments made by very 
large baseline interferometry(VLBI) \cite{Pyne}, we conclude that the external solution in this theory for 
those values of $\omega$ is the same as in the case of the superconducting cosmic
string in scalar-tensor theory \cite{Cris}. Thus, we will assume that the metric function
$\psi(r)$ has the same form obtained in the case without torsion\cite{Cris},and then, it can
be written as

\noindent

\begin{equation}
\psi(r)= {\frac{{(\frac{r}{r_0})}^n(1+p)}{(\frac{r}{r_0})^{2n} + p}},
\end{equation}

\noindent
where now the parameter $n$ is such that the following relation holds:
$n^2 = \kappa^{-1}(\phi)\lambda^2 + m^2 $, which is the same result obtained in\cite{VBCF}.

\vspace{.5 true cm}

Therefore, the external metric for the TCSCS, takes the form

\begin{equation}
ds^2 =  \left( \frac{r}{r_0} \right)^{-2n} W^2(r) \left[
\left( \frac{r}{r_0}\right)^{2m^2} (dr^2+ dz^2) +
B^2r^2d\theta^2 \right] - \left( \frac{r}{r_0} \right)^{2n}
\frac{1}{W^2(r)} dt^2 \label{m8},
\end{equation}

\noindent
with $W(r) = [(r/r_0)^{2n} + p]/[1+p]$. In order to get all informations concerning the current in the core
of the string we will use the the weak field approximation (For details of this procedure see Ref.\cite{Cris}),
in which case we can write the following relations

\begin{equation}
\begin{array}{ll}
g_{\mu\nu} = \eta_{\mu\nu} + h_{\mu\nu} ,\\
\Lambda(\phi) = \Lambda(\phi_0) +  \Lambda'(\phi_0)\phi_{(1)}, \\
T_{\mu \nu} = T_{(0)\mu\nu} + T_{(1)\mu \nu},\\
\phi = \phi_0 + \phi_{(1)},
\end{array}
\end{equation}

\noindent
where $\Lambda'(\phi_0) =\Lambda(\phi_0) \alpha(\phi_0)$,
$\eta_{\mu\nu} = diag(-,+,+,+)$ is the Minskowski metric
tensor and $\phi_0$ is a constant.

In this case, the energy-momentum tensor of the string source $ T_{(0)\mu
\nu}$ (in Cartesian coordinates)  has the following components

\begin{equation}
\begin{array}{ll}
T_{(0) tt} = U \delta(x)\delta(y) + \frac{Q^2
}{4\pi}\nabla^2 \left(ln\frac{r}{r_0}\right)^2 ,\\
T_{(0) zz} = -\tau \delta(x)\delta(y) + \frac{Q^2
}{4\pi}\nabla^2\left(ln\frac{r}{r_0}\right)^2,\\
T_{(0) ij} = -Q^2  \delta_{ij}\delta(x)\delta(y)
+ \frac{ Q^2  }{2 \pi}\partial_i \partial_j ln(r/r_0),
\end{array}
\end{equation}

\noindent
where the energy per unit length $U$ and the tension per unit length $\tau $, are given, respectively, by

\begin{equation}
U  = -2 \pi \int_0^{r_0} T^t_{_{(0)}t} r dr,
\end{equation}

\noindent
and

\begin{equation}
\tau = - 2\pi \int_0^{r_0} T^z_{_{(0)}z} r dr,
\end{equation}
\noindent

\noindent
and the charge in the core is given by

\begin{equation}
Q = - 2\pi e \int_0^{r_0} r dr \sigma^2 A_t.
\end{equation}
\noindent

Now, let us find the matching conditions connecting the internal and external solutions. For
this purpose, we shall use the linearized Einstein-Cartan equation in order to find the internal
solution. These equations are given by

\begin{equation} \nabla^2 h_{\mu\nu} =-16 \pi G (T_{(0)\mu \nu} -
\frac{1}{2} g_{_{(0)} \mu \nu} T_{(0)}).\label{ricci1}
\end{equation}

At this point let us use a method applied by Linet\cite{Linet} to solve the linearized Einstein's
equations using distribution functions. Doing this, we find that the internal solution
of Eq.(\ref{ricci1}) with time-independent source is given by

\begin{equation}
\begin{array}{ll}
h_{tt} =-4 \tilde G_0 [Q^2(\ln(r/r_0))^2 + (U-\tau -Q^2)\ln(r/r_0)] \\
h_{zz} =-4 \tilde G_0 [Q^2 \ln(r/r_0))^2 + ( U-\tau +Q^2 )\ln(r/r_0)] \\
h_{ij} = 2 \tilde G_0 [Q^2  r^2 \partial_i
\partial_j  - 2 \delta_{ij} (U+\tau - Q^2 ) \ln(r/r_0)].
\end{array}\label{agas}
\end{equation}

It is worth calling attention to the fact that torsion does not appear explicitly
in the components of the metric. This is a consequence of the linearized approximation in
the $\phi $-field.

In this case, we can find the matching
conditions using the fact that $[\{^\alpha_{\mu\nu}\}]_{_{r=r_0}}^{(+)}=
[\{^\alpha_{\mu\nu}\}]_{_{r=r_0}}^{(-)}$ and given by   $[g_{\alpha \rho}
K_{(\mu \nu)}^{\hspace{.3 true cm} \rho}]_{_{r=r_0}}^{(-)} = [g_{\alpha \rho} K_{(\mu \nu)}^{\hspace{.3 true cm}
\rho}]_{_{r=r_0}}^{(+)}$, for to contorsion\cite{Kopczynski,Volterra} where $(-)$ represents the internal
region and $(+)$ corresponds to the external region around $r = r_0$ .
Then the continuity conditions are

\begin{eqnarray}
&[g_{\mu \nu}]_{_{r=r_0}}^{(-)} =
[g_{\mu \nu}]_{_{r=r_0}}^{(+)}, \nonumber \\
&[\frac{\partial g_{\mu
\nu}}{\partial x^{\alpha }}]_{_{r=r_0}}^{(-)} =
[\frac{\partial g_{\mu \nu}}{\partial x^{\alpha}}]_{_{r=r_0}}^{(+)}
, \label{junc1}
\end{eqnarray}

Now, let us find the solution of the equation
for the field $\phi$. It is given by

\begin{equation}
\Box _g\phi_{(1)} =-4\pi \kappa^{-1} \tilde G_0\alpha (\phi )T_{(0)} \label{phi0},
\end{equation}

\noindent
where $ T_{(0)}= - (U + \tau + Q^2 ) \delta(x) \delta(y)  $ . Then, the solution
of Eq.(\ref{phi0}) in terms of the new coordinate, $\rho = r \left[ 1 + \tilde{G}_{0} (4\tau - Q^2) -
4\tilde{G}_{0} \tau \ln \frac{r}{r_0} +
2\tilde{G}_{0}Q^2 \ln^2 \frac{r}{r_0}\right] $,
is given by

\begin{equation}
\phi_{(1)} = 2\tilde{G}_{0} \kappa^{-1}(\phi) \alpha(\phi_0) (U +
\tau + Q^2 ) \ln \frac{\rho}{r_0}. \label{phi1}
\end{equation}

Taking into account the linearized forms of the exterior and interior metrics,
we obtain

\begin{eqnarray}
m^2 & = & 4\tilde{G}_0 Q^2 \nonumber \\
B^2 & = & 1 - 8\tilde{G}_0 (\tau - \frac{Q^2}{2}) \nonumber \\
\lambda & = & 2\tilde{G}_0 \alpha(\phi_0) (U+\tau+Q^2) \nonumber .
\end{eqnarray}

Finally, if we rewrite Eq.(\ref{agas}) in terms of the new coordinate, $\rho$,
and use the result given by Eq.(\ref{phi1}), the metric for a TCSCS, in the Einstein frame,
can be written as

\begin{eqnarray}
ds^2 & = & \left\{ 1 - 4\tilde{G}_0 \left[ Q^2\ln^2 \frac{\rho}{r_0} +
(U-\tau +Q^2)\ln
\frac{\rho}{r_0}\right] \right\} (d\rho^2 + dz^2) \nonumber \\
& & - \left\{ 1 + 4\tilde{G}_0 \left[ Q^2 \ln^2\frac{\rho}{r_0} +
(U-\tau -Q^2) \ln \frac{\rho}{r_0} \right]\right\} dt^2 \label{m5} \\
& & + \rho^2 \left[ 1 - 8\tilde{G}_0 (\tau- \frac{Q^2}{2}) -
4\tilde{G}_0 (U-\tau -Q^2)\ln \frac{\rho}{r_0} -
4\tilde{G}_0 Q^2 \ln^2\frac{\rho}{r_0} \right] d\theta^2 \nonumber .
\end{eqnarray}

Now, if we return to the Jordan-Fierz frame, the deficit angle associated with this space-time,
can be written in the linearized approximation as

\begin{equation}
\Delta\theta = 4\pi \tilde{G}_0 (U +\tau - 2Q^2) .\label{theta1}
\end{equation}

This means that in this order of approximation
there is no contribution arising from torsion. In fact, the contribution to the metric
due to torsion in the Jordan-Fierz frame appears in the dilaton solution given
by Eq.(\ref{phi1}), but it is not preserved in the linearized expression given by (\ref{theta1}), for the
deficit angle. The reason for this absence of torsion in the deficit angle is that the
contribution arising from torsion comes out only in second order in $\tilde{G}_0$ and therefore
it does not appear in the linearized solution we have considered. This result corresponds
to the same one obtained in the time-like case of pure scalar-tensor theories of gravity\cite{Oliveira}.

\section{Particle deflection near a TCSCS \zero }

In the previous section we concluded that torsion does not contribute to the deficit
angle, but some new physical effects appear associated with torsion in such a way that it
plays a role as we shall see in what follows. In this section we study the geodesic equation in the
space-time under consideration. To do this, we have to work with the metric given by Eq.(\ref{m5})
which was written in the Jordan-Fierz frame. For this we can use
the Christoffel symbols in Jordan-Fierz frame as in (\ref{Chisjf}).  In this frame, the tt-component
of Eq.(\ref{Chisjf}) is given by

\begin{equation}
\{^{i}_{tt } \}_{JF}  = \{^{i }_{tt } \} + K_{(t t)}^{\hspace{ .4 true cm} i }
 \label{Chisjf2},
\end{equation}

Let us consider the effect of the torsion on an uncharged particle moving around the defect,
assuming that the particle has a speed $|{\bf v}| \leq 1 $. In this case the geodesic
equations become

\begin{equation}
\frac{d^2x^i}{dt^2} + \left\{ ^{\hspace{.1 true cm} i}_{tt}\right \}_{JF} =0, \label{geo1}
\end{equation}

\noindent
where  $i$ is the spatial coordinate index. We note here that the simetric part of the dilaton 
gradient does not appear because the dilaton has no dependence on time. In the present
case, in the linearized approximation, the Christoffel symbols in the Einstein frame are
given by

\begin{equation}
{\left\{ ^{\hspace{.1 true cm} i}_{tt}\right\} }  = -\frac{1}{2} \partial^i h_{tt},
\end{equation}

\noindent
with $ g_{tt}= -1 +  h_{tt} $.
According to the previous section, in the context of
Einstein gravity the component $h_{tt}$ reads

\begin{equation}
h_{tt}=  -4\tilde G_0\left\{(U -\tau -Q^2)\ln(\frac{\rho}{ r_0}) + Q^2\ln(\frac{\rho}{r_0})^2
\right\}\label{heinstein}
\end{equation}

In order to make our analysis more simple, let us consider this approach in the framework of 
Einstein gravity. In this case, it is necessary to compute the symmetric part of the contortion  
(\ref{contorsion1}), which is given by

\begin{equation}
K^{\hspace{.2 true cm}r}_{(tt)} = -\frac{\tilde \phi'}{2\tilde
\phi}\sim \alpha(\phi_0) \phi_{(1)}'=
2\frac{1}{\rho}\tilde{G}_{0}\kappa^{-1}(\phi) \alpha^2(\phi_0) (U+\tau +Q^2).\label{contorcao2}
\end{equation}

\noindent
and the geodesic equations can be written as

\begin{equation}
\frac{d^2x^i}{dt^2} - \frac{1}{2} \partial^i  h_{tt} +
K_{(tt)}^{\hspace{.2 true cm} i}   =0, \label{geo4}
\end{equation}

\noindent
where $h_{tt}$ does not contain a contribution arising from torsion as we can see from
(\ref{heinstein}). We can see that this expression is compatible with the Christoffel
symbols in Jordan-Fierz frame calculated with the general form (\ref{Chisjf}).
Then, in this framework the contribution arising from torsion is contained entirely in the symmetric part
of the contorsion, $K_{(tt)}^{\hspace{.2 true cm} i}$, given by (\ref{contorcao2}), evaluated in the Jordan-Fierz
frame. This form of the geodesic equations is more suitable to an analysis of the new contribution arising
from the dilaton-torsion aspects.

Taking into account previous considerations we can conclude that the gravitational acceleration
induced by the string around it, is given by

\begin{equation}
a =  \nabla h_{tt} - 2\frac{\tilde{G}_{0} \kappa^{-1}(\phi)\alpha^2(\phi_0)
(U+\tau +Q^2)}{\rho},
\end{equation}

\noindent
and as a consequence, the torsion contribution to the force reads

\begin{equation}
f_{_{tors}}= -\frac{2 m}{\rho}\tilde{G}_{0}\kappa^{-1}(\phi) \alpha^2(\phi_0)
(U+\tau +Q^2)
\end{equation}

Therefore, the total force on a test particle due to a TCSCS can be written as

\begin{equation}
f = -\frac{4\tilde G_{0} m}{\rho} \left[ Q^2\left(\frac{(U-
\tau )}{Q^2}-1 + 2\ln(\rho/r_0) \right) +
\frac{1}{2}\kappa^{-1}(\phi)\alpha^2(\phi_0) Q^2(1+
\frac{(U+\tau)}{Q^2} )\right]. \label{force}
\end{equation}

In what follows, let us consider the deflection of particles moving past the string.
Assuming for simplicity that the direction of  propagation is perpendicular to the string,
we can write the metric, in terms of Minkowskian coordinates, as

\begin{equation}
ds^2 = (1 - \tilde h_{tt})[-dt^2 +dx^2 +dy^2]\label{lin2}
\end{equation}

\noindent
where $\tilde h_{tt}$ is given by

\begin{equation}
\tilde h_{tt} = -4\tilde G_0\left\{Q^2 (\ln(\frac{\rho}{r_0})) +  U -\tau -Q^2
 + \frac{\alpha^2 \kappa^{-1}(\phi)}{2}(U + \tau +
Q^2)\right\}\ln(\frac{\rho}{ r_0})\label{htt},
\end{equation}

Note that in this case, we can conclude that there is a change in the geodesics due to
the presence of the contortion (\ref{geo4}). In order to investigate the formation of a wake
moving behind a TCSCS, we will first consider the rest frame of the string with a velocity $v$
in the x direction. Thus, in this situation, the geodesic equations (\ref{geo4}) can be written in
the linearized version, as

\begin{equation}
2\ddot{x} = -(1-\dot{x}^2-\dot{y}^2)\partial_xh_{tt} +
(1-\dot{y}^2)\alpha(\phi_0)\partial_x\phi_{(1)}\label{y0},
\end{equation}

\begin{equation}
2\ddot{y} = -(1-\dot{x}^2-\dot{y}^2)\partial_yh_{tt} +
(1-\dot{x}^2)\alpha(\phi_0)\partial_y\phi_{(1)} \label{y},
\end{equation}

\noindent
where $h_{tt}$ is given by (\ref{heinstein}) and the overdot denotes derivative with respect to $t$.
Now, let us concentrate our attention in the last terms of Eqs.(\ref{y0}) and (\ref{y}) due to the fact
that they contain informations concerning the roles of the scalar field and torsion.
For our analysis it is enough to consider only terms of first order in $\tilde G_0$,
in  which case (\ref{y}) can be integrated over the unperturbed trajectory $x =vt$, $y=y_0$. Doing
this procedure and going to the frame in which the string has a velocity $v$, we find that
particles entering the wake have a transverse velocity given by

\begin{equation}
v_t = 4\pi \tilde G_0(U +\tau - 2Q^2)v \gamma +
\frac{4\pi \tilde G_0\left\{Q^2 (\ln(\frac{\rho}{r_0})) +  U -\tau -Q^2
 + \frac{1}{2}\alpha^2 \kappa^{-1}(\phi)(U + \tau + Q^2)\right\}}{v \gamma}\label{transversal}
\end{equation}

This result tell us that the first term contains the usual contribution of the deficit angle to the 
velocity of the particles. The second term contains the contributions arising from torsion and
electromagnetic field. A quick glance at this equation allow us to understand the
essential role played by torsion  in the context of the present
formalism. For example, if torsion is present, even in the case in which the
string has no current, an attractive gravitational force comes out.
In the context of the TCSCS, torsion enhances the force that a test particle feels outside
the string. This peculiar fact may have meaningful astrophysical as well as cosmological
implications, as for example, contributing to the process of formation of structures.

\section{Zel'dovich approximation in torsion space-time \zero}

\noindent

In this section we study the linear perturbation method developed by
Zel'dovich\cite{Zel} and apply it to the case of a space-time with torsion.
We will assume that dark matter particles interact very weakly, so that all forces on the particles of
non-gravitational origin can be ignored. In the case of the
ordinary cosmic strings this approach has been tested against exact solution and N-body
simulations with satisfactory results\cite{Bertschinger85}. An important advantage of the
Zel'dovich approach is the fact that it can be used in the case where the evolution is
strongly nonlinear. In this paper we consider a linear evolution. We analyse the
perturbation in the  dark matter background when the cosmic string is formed considering an
unperturbed universe. We also investigate the accretion of cold dark matter by straight
strings in a universe in the matter dominated era $t> t_{eq}$, with scale factor
$a(t) \sim t^{2/3}$, and average density given by

\begin{equation}
\rho_{av} = \frac{1}{6 \pi G t^2}.
\end{equation}

Let us write the trajectories of a cold dark matter particle as

\begin{equation}
\vec r(x,t) = a(t) [ \vec x + \Psi(x,t)], \label{dark1}
\end{equation}

\noindent
where $\vec x $ is the unperturbed comoving position of a particle
and $\psi $ is the comoving displacement measured from the position of the particle.

Now, let us analyse the equation of motion for this particle in presence of torsion.
We find a Newton's like equation in
Cartesian coordinates in the background corresponding to the metric (\ref{lin2}), which
reads as

\begin{equation}
\frac{d^2\vec r}{d t^2} = - \vec \nabla_r \Lambda
\end{equation}

\noindent
where the gravitational potential $\Lambda $ is given by

\begin{equation}
\Lambda = \Lambda_{CDM} + \Lambda_{TCSCS},
\end{equation}

\noindent
with $\Lambda_{CDM}$ being the gravitational potential due to the cold dark matter. In this
work we will consider that the presence of the cold dark matter does not affect the cosmic
string configuration, but on the other hand, the cosmic string perturbes the dark matter
trajectories. In this case, the  cosmic string
gravitational potential  $\Lambda_{TCSCS}$ can be written as a function of the
linearized cosmic string metric given by

\begin{equation}
\Lambda_{TCSCS} = -2\tilde G_0\left\{Q^2 (\ln(\frac{\rho}{r_0})) +  U -\tau -Q^2
 + \frac{\alpha^2 \kappa^{-1}(\phi)}{2}(U + \tau +
Q^2)\right\}\ln(\frac{\rho}{ r_0})\label{htt2},
\end{equation}
and therefore, the trajectories of the cold dark matter particles are perturbed by the
TCSCS space-time.

The gravitational potential $\Lambda(x,t)$ satisfies the Poisson equation

\begin{equation}
\nabla^2_r \Lambda = 4 \pi G(\rho + \rho_{tcscs})\label{nabla1},
\end{equation}

\noindent
where $\rho $ is the cold  dark matter density and $\rho_{tcscs}$ is the
perturbation due to the string. Mass conservation implies that

\begin{equation}
\rho(\vec r,t) = \frac{\rho_{av}(t) a^3(t)}{| det(\partial \vec r/\partial \vec x)|}.
\end{equation}

Using (\ref{dark1}) and (\ref{geo4}), thus we obtain

\begin{equation}
\frac{\partial \vec r}{\partial \vec x} = a(t)(1 +
\nabla_x . \Psi(\vec x, t)) \label{EC12},
\end{equation}

\noindent
and then, in the approximation we are considering, we get

\begin{equation}
\rho(\vec r, t) = \rho_{av}(t) (1 - \nabla_x . \Psi(\vec x, t)) \label{EC13}.
\end{equation}

>From Eq.(\ref{dark1}) for the trajectories of cold dark matter particles, we can obtain the
relations

\begin{equation}
\ddot r = \ddot a \Psi + 2 \dot a \dot \Psi + a \ddot \Psi \label{EC15}.
\end{equation}

\noindent
Now, consider the fact that $a \sim t^{2/3}$.Thus, we have

\begin{equation}
\ddot \Psi + \frac{4}{3t} \dot \Psi - \frac{2}{3t^2}\Psi =-
\frac{v_t}{a^3}\label{psi2}
\end{equation}

In order to solve Eq.(\ref{psi2}), let us consider an idealized situation in which
the cosmic string is formed in the time $t_i > t_{eq}$, in an initially
unpertubed universe. The perturbation caused by the cosmic string in the time
$t > t_i$ can be found by solving Eq.(\ref{psi2}) with the following initial conditions

\begin{equation}
\Psi(\vec x,t_i) = \dot \Psi (\vec x, t_i)=0.
\end{equation}

The solution of Eq.(\ref{psi2}), with the initial conditions given above, can be written as

\begin{equation}
\Psi(\vec x, t) = 3v_t
\left[1- \frac{2}{5}\frac{t_i}{t}-\frac{3}{5}\left(\frac{t}{t_i}\right)^{(2/3)}\right]
\end{equation}

\noindent
with $v_t$ given by (\ref{transversal}). This result shows the influence of
the parameters contained in the expression for $v_t$ given in E.(\ref{transversal}), which
determines the TCSCS space-time. It represents the time-dependent
accretion of the cold dark matter by a TCSCS. As we saw in the last Section the transversal
velocity of the accretion by wake, $v_t$ depends on the gravitational effects arising from torsion.
We can identify torsion through the presence of the parameter $\kappa(\phi) $.
Using Eq.(\ref{kappa}) we conclude that this contribution
is given by $ \alpha^2(1- 2 \epsilon \alpha^2)^{-1}\frac{(U + \tau + Q^2)}{v \gamma}$,
and therefore it enhances the atraction between the
cold dark matter particles in the wake and introduces new parameters that can be ajusted
by simulation in order to describe the observational spectrum.
This contribution comes in the contortion part given by (\ref{contorcao2}). The factor $\epsilon $, 
in the second term, contains torsion signature, which we are considering as an arbitrary parameter.
Nowdays, this parameter is taken to be very small, but in cosmic string scale formation or in
high energy scales\cite{Sengupta}, torsion effects  probably give us an interesting contribution.
In the case were $\epsilon $ can be neglected, the effect of torsion is to enhances dilaton effect.
Then, torsion effects are important and if analysed when $t \sim t_i$, it may be relevant and thus,
cannot be neglected. As we are working in the linearized approximation, the results obtained give
us only an approximate idea of the effects of torsion.

It is worth calling attention to the fact that in the present case, the time-dependent
solution has the same form of the ordinary cosmic string, but the accretion velocity
has a interesting dependence on the dilaton solution that contains  the current-carrying and
torsion terms.

\section{Wake evolution in a background with torsion  \zero}

In this section we study the formation of sheet-like wakes when a TCSCS moves fastly.
In the previous section we investigated the effect of cosmic string formation in the cold
dark matter universe, considering an unperturbed universe.
Now, we will make a quantitative description of accretion onto wakes using the Zel'dovich approximation
developed in last section, but now considering the presence of torsion and assuming that in the linearized
solution, torsion has some contribution. To investigate this problem, we have to solve the following
differential equation

\begin{equation}
\ddot \Psi + \frac{4}{3t} \dot \Psi - \frac{2}{3t^2}\Psi =-
\frac{u_p}{a^3}\label{psi3}
\end{equation}

\noindent
where $u_p$ is  the linearized contribution coming from the dilaton-torsion in the background under
consideration and is given by

\begin{equation}
u_p =
\frac{2\pi \tilde G_0 \kappa^{-1} \alpha^2(\phi_0)(U +
\tau + Q^2)}{v \gamma }\label{transversal1}
\end{equation}

In this case, differently from the last section, the wakes begin to move
with a velocity due to the perturbation, with magnitud given by (\ref{transversal1}).
We will consider that the wakes have a dissipation only induced by torsion
effects in the geodesic equation  and that the appropriate initial conditions are

\begin{equation}
\Psi(t_i)=0 \hspace{.3 true cm},  \dot \Psi(t_i)=-v_t
\end{equation}

\noindent
where $v_t$ contains the contribution of the torsion.
The solution in a time imediately after $t_i$, where the torsion effects are present,
is given by

\begin{equation}
\Psi(x,t) = \frac{3}{2}\left[ u_p- \frac{1}{5}(u_p- v_tt_i)\frac{t_i}{t} -
\frac{3}{5}(u_p-\frac{2}{3}v_tt_i)\left(\frac{t}{t_1}\right)^{2/3}\right] \label{ACC}
\end{equation}

Note that, when $t > > t_i$, the torsion perturbation $u_p$ can be neglected. In this situation,
there is a contribution coming from the torsion contained in $v_t$, which is small, but does
not vanish nowdays. Thus, taking into account this fact, Eq.(\ref{ACC}) turns into

\begin{equation}
\Psi(x,t)= \frac{3}{5}v_t\left[\frac{t_i^2}{t}- t_i\left(\frac{t}{t_i}\right)^{2/3}\right].
\end{equation}

The turn around surfaces, where particles stop expanding with the Hubble flow in the
x-direction and begin falling back towards the wake, can be found from the condition $\dot r_x =0 $ or,
equivalently, from $x +2 \Psi(x,t) =0$. This yields

\begin{equation}
x(t) = 3v_t\left[ \frac{u_p}{v_t} - \frac{1}{5}(\frac{u_p}{v_t}- t_i)\frac{t_i}{t} -
\frac{3}{5}(\frac{u_p}{v_t}-\frac{2}{3}t_i)\left(\frac{t}{t_i}\right)^{2/3}\right].
\end{equation}

The wake thickness $d(t)$ and the surface mass density $\sigma(t) $ of the wake are
given, respectively, by

\begin{equation}
 d(t) =2x(t)\left(\frac{t}{t_i}\right)^{2/3} \sim 6v_t\left[ \frac{u_p}{v_t} -
\frac{1}{5}(\frac{u_p}{v_t} -t_i)\frac{t_i}{t} -
\frac{3}{5}(\frac{u_p}{v_t} -\frac{2}{3}t_i)\left(\frac{t}{t_i}\right)^{2/3}
\right]\left(\frac{t}{t_i}\right)^{2/3},
\end{equation}

\begin{equation}
\sigma(t) = \rho_{t}d(t)= \sim \frac{v_t}{\pi \tilde G_0 t^2}\left[ \frac{u_p}{v_t} -
\frac{1}{5}(\frac{u_p}{v_t} - t_i)\frac{t_i}{t} -
\frac{3}{5}(\frac{u_p}{v_t} -
\frac{2}{3}t_i)\left(\frac{t}{t_i}\right)^{2/3} \right]\left(\frac{t}{t_i}\right)^{2/3}.
\end{equation}

We investigated the wake evolution in a background with torsion using
$\rho(t)= \frac{1}{6\pi \tilde G_0t^2}$. This is the same expression as in the case of the flat
universe, and constitutes a reasonable approximation in the case of a linearized solution in terms
of torsion in the matter-dominated era with the wake formed at $t_i \sim t_{eq}$.
The most important new feature of the results here presented is to consider the wake in a
background with torsion.

In the wake evolution case the contribution of the background to the
wake vanishes and the dilaton-torsion contribution only appears in
the transversal velocity of the accretion, $v_t $. In the case where the torsion
parameter $\epsilon $ is small, this contribution can  be neglected as compared with the electromagnetic
effects and the only effect of the torsion is to amplify the dilaton interactions.
But if we analyse the result when $t\sim t_i$, we conclude that the terms which contains
contributions arising from torsion can be relevant and the $\epsilon$ parameter can be ajusted 
in order to corresponds to the early era. In this scenario, torsion could dominate the accretion of matter.
Other interesting analyse can be done when the electromagnetic current vanishes, in which case the contribution 
is due only to dilaton-torsion effects, in which case $\Psi(x,t)$ is given by 
$\Psi(x,t)= \frac{3}{5}u_p\left[\frac{t_i^2}{t}- t_i\left(\frac{t}{t_i}\right)^{2/3}\right].
$ In this scenario, at high energy scale, these effects can be measurable\cite{Sengupta}, in principle.

\section{Conclusion}

We have obtained the solution that corresponds to a time-like current-carrying screwed
cosmic string (TCSCS). Screwed cosmic strings are stable topological defects and has been obtained
in the framework of a general scalar-tensor theory including torsion. In the model in which
spin vanishes, torsion is a $\phi$-gradient and propagates outside the string. In fact,
torsion is small but gives a non-negligible contribution to the geodesic equations obtained from
the contortion term and from the scalar fields. The motivation to consider this scenario comes from
the fact that scalar-tensor gravitational fields are important for a consistent description of
gravity, at least at sufficiently high energy scales. On the other hand, torsion can induces
some physical effects and could be important at some energy scale, as for example, in the low-energy
limit of a string theory.

The analyse of the metric and contortion help us to understand the consequences
of the gravitational interaction due to a TCSCS at a cosmological level.
One important consequence is related with the gravitational field surrounding
a TCSCS, which is divergent for the state parameter
approaching the mass of the current carrier, and thus the gravitational effects
seem unbounded. However, it is important to call attention to the fact that this divergence
is strongly connected with another divergence, namely, that associated with the string tension: as U
increases, the tension decreases to zero, and eventually becomes negative so that
the corresponding state is absolutely unstable against the transverse perturbations
and should go into a stable state. Therefore, the gravitational effects of such
strings are indeed limited even at classical level.

In the space-time generated by a TCSCS, massless particles (such as photons) will be
deflected by an angle $\Delta \theta = 4 \pi \tilde G_{0}(U+
\tau - 2Q^2)$. From the observational point of view, it would be
impossible to distinguish a screwed string from its general
relativity partner, just by considering effects based on deflection
of light, as for instance, double image effect. On the other
hand, trajectories of massive particles will be affected by
torsion coupling, which is generated by a space-time with torsion.
\cite{Cris1,Kleinert2000}.

We have shown that wakes produced by the string  in one Hubble time
can have important effects due to torsion.
If the string is moving with normal velocity, $v$, through matter,
a transversal velocity appears.
It is worth calling attention to the fact that there exists, in this case,
a new contribution to the transversal velocity given by
$v_t = \frac{2\pi\tilde G_0 \kappa^{-1} \alpha^2(\phi_0)(U +
\tau + Q^2)}{v \gamma}$ which is associated with aspects of the scalar-tensor theories
which includes torsion.

We also have shown that the propagation of photons is unaffected by a TCSCS and
it is only affected by the angular deficit. This result shows us that the effect of torsion on massive
particles is qualitatively different from its effect on radiation. This
aspect becomes  especially relevant when calculating CMBR-anisotropy and
the power spectrum as wiggly cosmic strings.

The investigations concerning the formation and evolution of wakes in the space-time of TCSCS 
shows that there is an effect arising from torsion on the process of wakes formation. Using the
Zel'dovich approximation we analysed the linear perturbations in this space-time.
The accretion of cold dark matter in the isolated strength
cosmic string is studied and the effect of torsion was pointed out.
Assuming the validity of the linear perturbation methods developed by
Zel'dovich in this background and that dark matter particles interacts very weakly, in such a way that 
all forces different from the gravitational one can be ignored, it was shown that the accretion of matter by wakes
formation when a TCSCS moves with speed $v $ depends on the features of the scalar field and torsion.

Therefore, assuming that torsion have had a physically relevant role during the early
stages of the Universe's evolution, we can say that torsion fields may
be potentially sources of dynamical stresses which, when coupled to other
fundamental fields (i.e., the gravitational and scalar fields), might have
performed an important action during the phase transitions leading to
formation of topological defects such as the TCSCS we have considered.
Therefore, it seems an important issue to
investigate basic models and scenarios involving cosmic defects within the
context of scalar-tensor theories with torsion and one of the reasons for this is the fact that
torsion would be relevant in $t \sim t_i$, that is, in the early stages of our Universe.

\vspace{.5 true cm}

{\bf Acknowledgments:} We would like to express our deep gratitude to
Prof. J. A. Helay\"el-Neto for helpful discussions on this subject.
VBB and CNF would like to thank CNPq(Brazil) for financial support. AARS would like to thank
CAPES(Brazil) for a fellowship. VBB and AARS would like to thank
CAPES/PROCAD. We also thank Centro Brasileiro de Pesquisas F\'{\i}sicas (CBPF).


\begin{thebibliography}{99}

\bibitem{Jordan} M. Fierz, Helv. Phys. Acta {\bf 29}, 129 (1956); P. Jordan,
Z. Phys. {\bf 157}, 112 (1959); C. Brans and R. H. Dicke, Phys. Rev. {\bf 124}, 925 (1961).

\bibitem{Bergmann} P. G. Bergmann, Int. J. Theor. Phys. {\bf 1}, 25 (1968); K. Nordtvedt,
Astrophys. J. {\bf 161}, 1059 (1970); R. V. Wagoner, Phys. Rev. {\bf D1}, 3209 (1970).

\bibitem{Esposito} T. Damour and G. Esposito-Far\`ese, Class. Quantum Grav. {\bf 9}, 2093 (1992);
T. Damour and K. Nordtvedt, Phys. Rev. {\bf D48}, 3436 (1993).

\bibitem{Shapiro1} I. Buchbinder, S.D. Odintsov and I.Shapiro
{ \it Effective Action in Quantum Gravity}, ( Institute of Physics Publishing, Bristol and Philadelphia, 1992).

\bibitem{Gaspperini} V. de Sabbata and M.Gasperini, { \it Introduction to
Gravitation}, ( World Scientific Publishing, Singapore, 1985).

\bibitem{Hehl} F. W. Hehl et al., Rev. Mod. Phys. {\bf 48}, 393 (1976). 

\bibitem{Palle} D. Palle, Nuovo Cim. {\bf B114}, 853 (1999). 

\bibitem{Kim} S. W. Kim, Phys. Rev. {\bf D34}, 1011 (1986). 

\bibitem{will94} C. M. Will, Phys. Rev. {\bf D50}, 6058 (1994). 

\bibitem{will93} C. M. Will, {\it Experimental Tests of Gravity Theories}, revised ed.
( Cambridge University Press, Cambridge, England, 1993). 

\bibitem{dam} T. Damour, Nucl. Phys. ( Proc. Suppl.) {\bf 80}, 41 (2000). 

\bibitem{Vilenkin} A. Vilenkin, Phys. Rev. {\bf D 23}, 852 (1981);
W. A. Hiscock, Phys. Rev. {\bf D 31}, 3288 (1985); D. Garfinkel, Phys. Rev. {\bf D 32} 1323 (1985). 

\bibitem{Gott} R. Gott, Astrophys. J. {\bf 288}, 422 (1985). 

\bibitem{Aliev} A. N. Aliev and D. V. Gal'tsov, Ann. Phys. (N.Y.) {\bf 193}, 165 (1989).

\bibitem{Blinet} B. Linet, Phys. Rev. {\bf D33}, 1833 (1986).

\bibitem{Ford} L. H. Ford and A. Vilenkin, J. Phys. {\bf A14}, 2353 (1981); V. B. Bezerra,
Phys. Rev. {\bf D35}, 2031 (1987); id. Ann. Phys.(N.Y.), {\bf 203 }, 392 (1990).

\bibitem{rom} C. Gundlach and M. E. Ortiz, Phys. Rev. {\bf D42}, 2521 (1990);
L. O. Pimentel and A. No\'e Morales, Revista Mexicana de F\'{\i}sica {\bf 36}
S199, (1990); A. Barros and C. Romero, J. Math. Phys. {\bf 36}, 5800 (1990). 

\bibitem{greg} R. Gregory and C. Santos, Phys. Rev. {\bf D56}, 1194 (1997). 

\bibitem{mexg} M. E. X. Guimar\~aes, Class. Quantum Grav. {\bf 14}, 435 (1997);
 R. M. Teixeira Filho and V. B. Bezerra, Phys. Rev. {\bf D64}, 067502 (2001);
V. B. Bezerra, L. P. Colatto, M. E. X. Guimar\~aes and R. M. Teixeira Filho,
Phys. Rev. {\bf D65}, 104027 (2002). 

\bibitem{Cris} C.N.Ferreira, M.E.X. Guimar\~aes and
J.A.Helayel-Neto, Nucl.Phys. {\bf B 581}, 165 (2000). 

\bibitem{William} William M. Baker, Class. Quantum Grav. {\bf 7}, 717 (1990). 

\bibitem{Adak} M. Adak, T. Dureli and L. H. Ryder, Class. Quantum Grav. 
 {\bf 18}, 1503 (2001).

\bibitem{Sabbata1} V. De Sabbata, IL Nuovo Cimento, {\bf 107 A}, 363 (1994). 

\bibitem{Yishi} Y.Duan, G. Yang and Y. Jiang, Helv. Phys. Acta
 {\bf 70}, 565 (1997). 

\bibitem{Trautman} A. Trautman, Nature {\bf 242}, 7 (1973). 

\bibitem{Stewart} J. Stewart and P. Hajicek, Nature
{\bf 244}, 96 (1973). 


\bibitem{Niew} P. Van Nieuwenhuizen, Phys.Rep. {\bf 68}, 189 (1981). 

\bibitem{Demianski} M. Demianski, R. De Ritis, G. Platania,
P. Scudellaro and C. Stornaiolo, Phys. Rev {\bf D 35}, 1181 (1987). 

\bibitem{Ellis} G. F. Ellis and M. Bruni, Phys. Rev. {\bf D 40}, 1804
(1989); G.F. Ellis and J. Hwang, Phys. Rev {\bf D 40}, 1819 (1989);
G. F. Ellis, M. Bruni and J. Hwang, Phys. Rev {\bf D 42}, 1035 (1990).

\bibitem{VBCF} V. B. Bezerra and C. N. Ferreira, Phys. Rev. {\bf D65}, 084030 (2002).

\bibitem{VBHMCF} V. B. Bezerra, H. J. Mosquera Cuesta and C.N. Ferreira, Phys.Rev. {\bf D67}, 084011, (2003).

\bibitem{Bennett88}  D. P. Bennett and F. R. Bouchet Phys. Rev . Lett. {\bf  60}, 257 (1988).

\bibitem{Bennett90}  D. P. Bennett and F. R. Bouchet Phys. Rev.  {\bf  D 41},  2408,  (1990).

\bibitem{Allen90}  B. Allen and E. P. S. Shellard, Phys. Rev. Lett. {\bf 64}, 119 (1990).

\bibitem{Albrecht89} A. Albrecht and N. Turok, Phys. Rev {\bf D 40}, 973 (1989).

\bibitem{Silk84} J. Silk and A. Vilenkin, Phys. Rev. Lett {\bf 53}, 1700 (1985).

\bibitem{Stebbins89} A. Stebbins, A. Veeraraghavan, R. H. Brandenberger,
J. Silk and N. Turok, Astrophys. J. {\bf 322}, 1 (1989).

\bibitem{Pogosian1} L. Pogosian and T. Vachaspati, Phys. Rev {\bf D 60} , 083504, (1990).

\bibitem{Pogosian2} L. Pogosian, Int. J. Mod. Phys. {\bf A 16SIC}, 1043 (2001).

\bibitem{Zel} Ya. B. Zel'dovich, Astron. Ap. {\bf 5}, 84 (1970). 

\bibitem{Linet} B. Linet, Class. Quantum Grav. {\bf 6}, 435 (1989).

\bibitem{Nielsen} H.B Nielsen and P. Olesen, Nucl.Phys. {\bf 61}, 45 (1973). 

\bibitem{PP} P. Peter and D. Puy, Phys. ev. {\bf D48}, 5546 (1993).

\bibitem{MacC} D. Kramer, H. Stephani, E. Herlt and M. MacCallum, {\it Exact solutions of Einstein's
field equations}, Cambridge University Press, cambridge, 1980). 

\bibitem{Pyne} T. Pyne et al., Astrophys. J. {\bf 465}, 566 (1996).

\bibitem{Kopczynski} W. Arkuszewski {\it et al.}, Commun. Math. Phys. {\bf 45}, 183 (1975). 

\bibitem{Volterra} R. A. Puntigam and H. H. Soleng, Class. Quantum Grav. {\bf 14}, 1129 (1997). 

\bibitem{Oliveira} A. L. N. Oliveira and M. E. X. Guimarães, '' Wakes in dilatonic
current-carrying cosmic strings'', hep-th/0303112 ( To appear in Pysical Review D).

\bibitem{Bertschinger85} E. Bertschinger, Astrophys. J. Suppl, {\bf 58}, 39, (1985).

\bibitem{Sengupta} B. Mukhopadhyaya, S. Sen and S. SenGupta, Phys.Rev.Lett. 89,  121101, (2002).

\bibitem{Cris1} C.N.Ferreira, Class. Quantum Grav., {\bf 19}, 741, (2001).

\bibitem{Kleinert2000} H.Kleinert, Phys.Lett.{ \bf B440} 283 (1998);
H. Kleinert, Gen. Rel. Grav. { \bf 32} 769 (2000); { \bf 32}, 1271 (2000).

\end{thebibliography}
\end{document}